# Griffiths phase and symmetry breaking in the hidden-order phase of $URu_2Si_2$

Yi Liu, Wen Zhang, Xiaoying Wang, Donghua Xie, and Xinchun Lai

Science and Technology on Surface Physics and Chemistry Laboratory, Huafeng Xincun 9, Jiangyou 621908, China

The heavy-fermion compound $URu_2Si_2$ exhibits a hidden-order phase below the temperature, ~ 17.5 K. In spite of intense research for past three decades, no consensus on the order parameter exists and the nature has posed a long-standing mystery. Here we report the discovery of a Griffiths phase within the hidden-order phase, characterized by residual short-range correlations on the collapse of long-range orders due to the dilution effects. In the Griffiths phase scenario, strong evidence are provided for those cluster-like spins, such as the unique power-law behavior of magnetic susceptibility and specific heat as well as the frequency dispersion of AC susceptibility. In this way, the existence of an order parameter is excluded, and the hidden order has a significant kinship with the long-range large-moment antiferromagnetism which is accessible by tuning the hydrostatic pressure or the chemical pressure (i.e., isoelectronic Fe doping). Moreover, an unidirectional anisotropy of resistivity measurements in rotating magnetic fields is observed in the hidden-order phase. The anisotropic magnetoresistance and the associated broken symmetries directly reflect the freezing behavior of at least part of magnetic clusters. Thus, the demonstrations of the Griffiths phase as an alternative proposal for the hidden-order phase of $URu_2Si_2$ are very promising, challenging the understanding of exotic electronic states in correlated matter and quantum materials.

In strongly correlated electron systems, a prime example of emergent phenomena is the hidden-order (HO) phase in the heavy-fermion compound URu$_2$Si$_2$, which has attracted an enormous amount of attention for over past three decades [1-6]. The HO phase occurs below $T_{HO}$ = 17.5 K and coexists with an unconventional superconductivity below $T_S$ = 1.5 K. In the HO phase, neutron-scattering experiments reveal a small antiferromagnetic moment of only 0.03 μ$_B$/U parallel to the tetragonal c axis [7], which is too small to account for the entropy of ~0.2Rln(2) derived from the specific heat anomaly [3]. Further, the small antiferromagnetic moment is thought to be extrinsic due to the presence of small antiferromagnetic regions induced by inhomogeneous strain [8, 9]. Many candidates for the order parameter have been proposed [5], however, the nature of the HO phase has not been definitively established in spite of some possibilities recently suggested [10-15].

The most important issue for URu$_2$Si$_2$ is that it exists in close proximity to a pressure-induced large-moment antiferromagnetism (LMAF), as shown in Fig. 1(a). The boundary between HO and LMAF phases is a first-order phase transition at a critical pressure that lies in the range 0.5-1.5 GPa [16-19]. Although order parameters are presumably different in HO and LMAF phases, the two phases exhibit nearly indistinguishable transport and thermodynamic properties [20]. It is thus believed that HO and LMAF phases are intimately related and a comprehensive understanding of both phases will be crucial in unraveling the nature of the HO phase. Recently, substitution of Fe for Ru provides an opportunity to probe HO and LMAF phases simultaneously at ambient pressure, since angle-resolved photoemission spectroscopy, scanning tunneling microscopy, and quantum oscillations are probably impossible to carry out at high pressures. As Ru ions are partly replaced by smaller Fe ions, the substitution acts actually as a chemical pressure and well reproduces general features of both phases in the phase diagram [21-27]. In this letter, we propose that the HO phase of $URu_2Si_2$ is identified as a Griffiths phase where the relation between HO and LMAF phases can be qualitatively described by the concept of the Griffiths phase in diluted Ising ferromagnets [28]. Our results clearly show the Griffiths phase emerging within the HO phase, which manifests itself by residual short-range correlations on the collapse of the long-range antiferromagnetic order.

High-quality single crystals of $URu_2Si_2$ were grown by the Czochralski pulling method in a tetra-arc furnace under protective argon atmosphere. Small oriented cubic lumps were cut from the cleaved ingots, and subsequently annealed at 1073 K under ultrahigh vacuum for a week. The quality of the samples was verified by means of X-ray diffractions. The residual resistivity ratio, RRR = $\rho(300\ K) / \rho(2\ K)$, of a relatively large lump amounts to 54, and in some tiny samples can reach a value greater than 600, confirming the high quality of the single crystals. Magnetic measurements were performed using a Quantum Design physical property measurement system (PPMS). Both DC and AC magnetic fields in measurements were applied along the tetragonal c direction, the easy axis of magnetization. In AC susceptibility measurements, an excitation of 0.8 mT was used, operating at 100 Hz, 300 Hz, 500 Hz, 700 Hz, 900 Hz, and 1300 Hz, respectively. In specific heat measurements, the thermal relaxation method on the PPMS platform was used, and Apiezon N vacuum grease was used as the sample mounting medium. In resistivity measurements in rotating magnetic fields, a standard four-wire AC technique was used with a PPMS rotator option. The resistivity was measured when the magnetic field of $\mu_0 H = 4$ T was rotated spanning the ac plane with high alignment precision.

Fig. 1(b) schematically shows the phase diagram of the Griffiths phase predicted in diluted Ising ferromagnets more than 40 years ago [28]. In a diluted Ising model with spins randomly located on the regular lattice, only a fraction of the lattice sites are occupied by Ising spins whereas the rest remains vacant. As a consequence, the Curie point ($T_C$) of the system varies with the occupation fraction (f), and no spontaneous magnetization occurs at a finite temperature with f less than the critical percolation probability, $f_C$. The Griffiths phase is denoted as the region below the dashed line of the initial $T_C$, in the absence of any long-range orders. Notably, the Griffiths phase can be considered as a paramagnetic phase coexisting with rare coupled magnetic clusters which have large susceptibilities [29]. The critical fraction points out a percolation effect at which an infinite cluster first appears as f increases. In real materials, the fact that f is independent of temperature, pressure, and other control parameters is not true, making the phase diagram more complicated than the theory prediction.

In the HO phase of $URu_2Si_2$, we demonstrate the existence of the Griffiths phase by conventional physical property measurements. Fig. 2(a) shows the temperature dependences of magnetic susceptibility, $\chi_{DC}$, measured at various magnetic fields. The $\chi_{DC}$ curves are plotted with vertical offset of finite values for clarity. Fig. 2(b) shows the electronic specific heat divided by the temperature, $C_e/T$. The electronic specific heat is determined by subtracting the phonon contribution from the total value, as discussed in a previous work [21]. Both figures emphasize on experimental data below $T_{HO}$, and are plotted in double-log scales for fits of the power-law functions. Surprisingly, all the low-temperature data follow the expression, $\chi \sim C/T \sim T^{-1+\lambda}$, with $0 < \lambda < 1$ predicted by the Griffiths phase [29]. In actual fact, our results in magnetic susceptibility and specific heat measurements are fully in accordance with previous reports on high-quality single crystals [21, 25, 30], but no attention has ever been paid to the power-law behavior of the Griffiths phase. Especially, in $^{29}Si$ NMR spectra measurements the temperature dependence of the Knight shift shows a similar behavior below $T_{HO}$ [31], unambiguously favoring the Griffiths phase scenario. In other heavy-fermion compounds, such as $Th_{1-x}U_xPd_2Al_3$, $Y_{1-x}U_xPd_3$, and $UCu_{5-x}M_x$ (M = Pd, Pt), measurements can likewise be analyzed with the context of the Griffiths phase, which may open an unified physical picture in a wide variety of quantum materials [32]. In the framework of quantum phase transitions in metallic ferromagnets, Griffiths-phase effects usually coexist with singularities close to the quantum critical points, and are even possible in clean undoped stoichiometric systems that should be less disordered [33]. For example, the ground state of CeFePO is a short-range ordered state which is close to a ferromagnetic instability [34]. In the Griffiths phase, a short-range order occurs only when the interaction of magnetic clusters is strong enough; conversely, if the interaction is weak a short-range order is not necessary.

In $URu_2Si_2$, a localized or itinerant description of the Ising-like magnetism is controversial on account of the dual nature of 5f electrons. Since most of the experiments suggest an itinerant magnetism, the conclusion on the sparsity of the unhybridized localized 5f moments is obvious. Nevertheless, if a completely itinerant magnetism is considered, it is not compatible with the evidence for a localized 5f character in quantum oscillations [35]. As a consequence, a phenomenological two-fluid model in which heavy itinerant quasiparticles (so called the Kondo liquid) coexists with unhybridized localized 5f electrons is essential [36]. The model can well describe both localized and itinerant 5f states, which may be crucial in understanding the microscopic origin of heavy-fermion physics. Additionally, because a mere remnant of unhybridized localized 5f moments is not sufficient to explain the anomalies at the HO transition, both two components of the two-fluid model have to be considered. Despite the sparsity of unhybridized localized 5f moments, they coexist with itinerant quasiparticles in the Griffiths phase of the HO phase as well as in the long-range antiferromagnetic order of the LMAF phase. When the system enters the Griffiths region, it shows sudden alterations in both the hybridization at each U atom and the associated heavy-fermion states. Simultaneously, a reconstruction of at least a section of the Fermi surface with a large entropy change is likely to happen.

To gain an insight into the HO phase, we plot the fitted parameter, $\lambda$, of the Griffiths phase as a function of the magnetic field. The value of $\lambda$ is close to 1, indicating that the coupling among magnetic clusters is rather weak, in close proximity to the temperature-independent paramagnetism. As the magnetic field increases, the power-law behavior in both magnetic susceptibility and specific heat measurements is likely to be smeared out where $\lambda$ shows a similar tendency of the saturation at around 1 T. In addition to the similarity of the field dependence, $\lambda$ of the specific heat is moderately smaller than that of the magnetic susceptibility. The difference in the response to external magnetic fields seems to point out the contribution of itinerant quasiparticles, distinguishing the two components of the two-fluid model. As revealed by the specific heat, itinerant quasiparticles carry an entropy, but are not sensitively detected by the magnetic susceptibility.

In the Griffiths phase scenario, the emergence of coupled magnetic clusters below $T_{HO}$ should be viewed as hidden-order signatures of $URu_2Si_2$, as detected by AC susceptibility, $\chi_{AC}$. Fig. 4(a) shows the temperature dependences of AC susceptibility measured at various frequencies, in the range from 100 Hz to 1300 Hz. All data show a kink at $T_{HO}$, and coincide exactly at temperatures above $T_{HO}$. Nevertheless, at temperatures below $T_{HO}$, AC susceptibility exhibits a clear frequency dispersion. As the frequency increases, the hump-like feature in the temperature dependence of AC susceptibility weakens and slightly shifts to higher temperatures, signifying the collective freezing of magnetic clusters. Fig. 4(b) shows the temperature dependences of AC susceptibility measured at various DC bias magnetic fields. At small bias fields, the hump of AC susceptibility below $T_{HO}$ drops rapidly, indicating the suppression of short-range correlations. When the bias field is strong enough, AC susceptibility shows an upturn and increases steadily on the contrary. In Fig. 4(c), we plot the relative change of AC susceptibility, $[\chi_{AC}(B)- \chi_{AC}(0)] / \chi_{AC}(0) \times 100\%$, as a function of the magnetic field. At high magnetic fields, AC susceptibility at low temperatures, and even at appreciably higher temperatures above $T_{HO}$, shows a gradual increase with increasing field. Such an increase is abnormal, which is attributed to the breakdown of part of the Kondo screening effects, resulting in an increase of the effective magnetic moment. Otherwise at 200 K, far higher than the coherence temperature of the crossover at 50~100 K, AC susceptibility shows a continuous monotonous decrease as the magnetic field increases.

Finally we comment on short-range characters of the spin dynamics and possible broken symmetries in the Griffiths phase. In recent years, two-fold oscillations or electronic nematic phases are observed by magnetic torque [12], elastoresistance [13], Raman spectroscopy [14, 15], and high resolution X-ray diffraction [38]. In view of the Griffiths phase, the broken rotational symmetry within the basal c plane is questioned, especially in tiny and ultra-pure crystals. If the coherence length is comparable with or even larger than the probe size, the symmetry breaking of short-range correlations has no physics. Apparently, this case is attributed to the asymmetrical distribution of magnetic clusters, and is supported by the polar Kerr effect measurements [37]. The Kerr rotation is measured in zero fields, but on entering the HO phase it shows a tiny remanence which sensitively depends on the magnitude of the training field and the field-cooling history. Actually, the remanent Kerr rotation appears at ~25 K, higher than $T_N$ of any antiferromagnetic transitions at pressures up to 2 GPa. Such a tiny non-zero magnetic moment may be intrinsic to the HO phase, and in this manner, a small fraction of magnetic clusters even persists at elevated temperatures. For a better description of these frozen cluster-like spins, we have performed high-sensitivity resistivity measurements in rotating magnetic fields, reflecting the magnetic anisotropy and the symmetry breaking. The resistivity, $\rho$, is measured when the magnetic field is varied within the ac plane as a function of the azimuthal angle, $\theta$. The details of resistivity measurements are described in supplementary materials. In Fig. 5, we show the temperature evolution of the relative change of the resistivity, $\rho(\theta)/\rho(\theta=0)$, at $\mu_0 H = 4$ T. At 20 K (in the paramagnetic phase), the curve is perfectly sinusoidal, showing a two-fold oscillations of the Ising-like magnetism along the tetragonal c axis. Whereas at temperatures below $T_{HO}$, an unidirectional anisotropy is distinctly observed in the HO phase, practically exceeding 8% at 5 K. The unidirectional anisotropy naturally breaks the two-fold rotational symmetry of the ac plane where the tetragonal c direction is parallel with the easy axis of magnetization. Meanwhile, both the unidirectional anisotropy and the remanent Kerr rotation provide direct arguments for the broken time-reversal symmetry. Thus, the symmetry breaking is in general a hallmark of electronic short-range correlations, unambiguously favoring the applicability of the Griffiths phase.

In summary, we show the establishment of the Griffiths phase within the HO phase of URu$_2$Si$_2$, as revealed by magnetic and specific heat measurements. The power-law behavior of the Griffiths phase, the frequency dispersion of short-range correlations, and the unidirectional anisotropy of the magnetoresistance are clearly observed in the

HO phase. In the Griffiths phase scenario, the long-range antiferromagnetic order disappears in the HO phase due to the dilution effects, displaying a significant kinship between HO and pressure-induced LMAF phases. Therefore, the HO phase has no order parameters, yielding just a percolation effect with or without the long-range order. In a phenomenological two-fluid model, itinerant quasiparticles should indeed make no much difference in HO and LMAF phases, which naturally explains why the Fermi surface practically shows no change across the HO-LMAF phase boundary [20]. The similarity of the Fermi surface is further supported by recently reported magnetic excitation spectrum, where gapped excitations at $\Sigma = (1.4, 0, 0)$ and $Z = (1, 0, 0)$ points in the HO phase persist in the LMAF phase, albeit with an enhanced gap at the $\Sigma$ point from 4.2 meV (0 GPa) to 5.5 meV (1.02 GPa) [39]. As expected, such excitations of itinerant quasiparticles are observed in all Fe-doped $U(Ru_{1-x}Fe_x)_2Si_2$ samples with $0.01 \leq x \leq 0.15$ in which the incommensurate gap increases continuously with Fe doping [27]. The gapping of these excitations has been proved to result in an entropy change of sufficient size to account for the specific heat jump at the HO transition [40].

Owing to the same reason, we conjecture that the superconductivity has not to be completely restricted within the HO phase. That is to say, the superconductivity may really coexist with the LMAF phase, as supported by resistivity measurements where the superconductivity even survives above 1.8 GPa [17, 41]. However, the detection of $T_S$ in resistivity measurements is queried in the case of sample inhomogeneities. One has to look at the specific heat to learn bulk signals of the superconductivity. Since in hitherto measurements the bulk anomaly at $T_S$ seems to just vanish on entering the LMAF phase, whether the superconductivity is exactly excluded by the LMAF phase remains an open question.

We thank Prof. Tao Wu, Dr. Qiuyun Chen, and Dr. Yun Zhang for helpful discussions. We acknowledge the support from the National Natural Science Foundation of China under Award 11404297, the Science Challenge Project under Award TZ2016004, the National Key Research and Development Program of China under Award 2017YFA0303104, and the Science and Technology Foundation of China Academy of Engineering Physics under Awards 2013B0301050 and 2014A0301013.

FIG. 1. Schematic phase diagrams of URu$_2$Si$_2$ (a) and diluted Ising ferromagnets (b). The hidden order (HO) and its adjacent large-moment antiferromagnetism (LMAF) are well described by the context of the Griffiths phase (GP) predicted in Ising-like ferromagnetism (FM) due to the dilution effects.

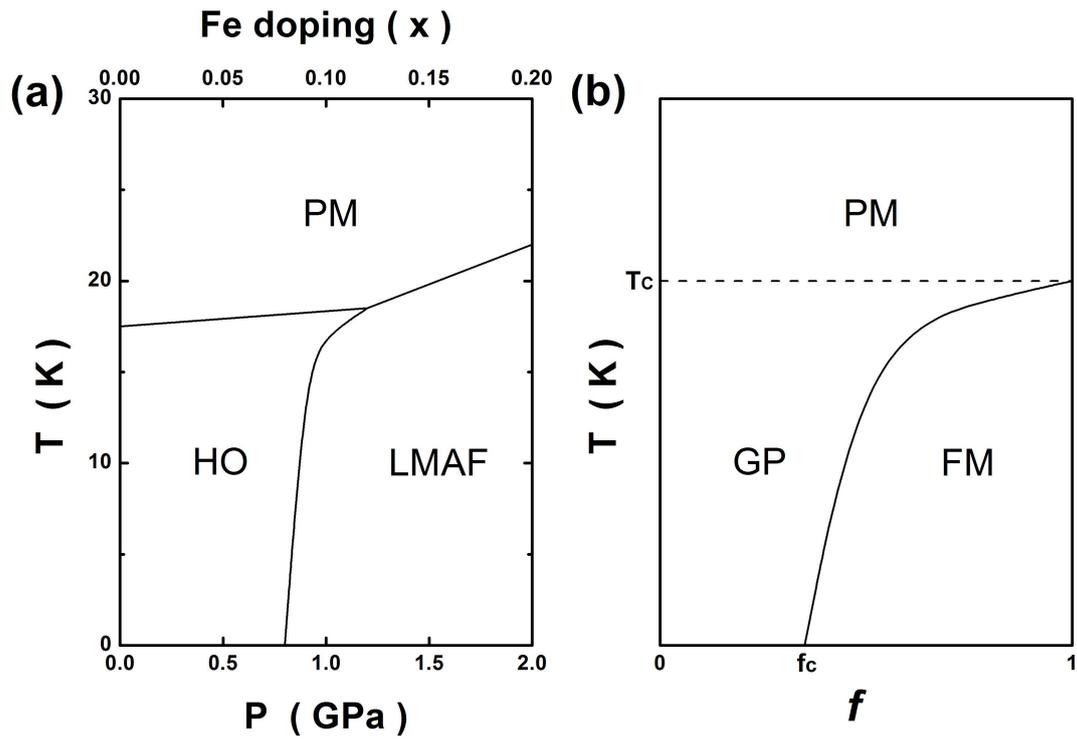

FIG. 2. Temperature dependences of magnetic susceptibility (a) and electronic specific heat (b) of $URu_2Si_2$. The specific heat is shown in C/T versus T plots. Both figures are shown in double-log scales for fits to the power-law function of the Griffiths phase, $\chi \sim C/T \sim T^{-1+\lambda}$, with $0 < \lambda < 1$.

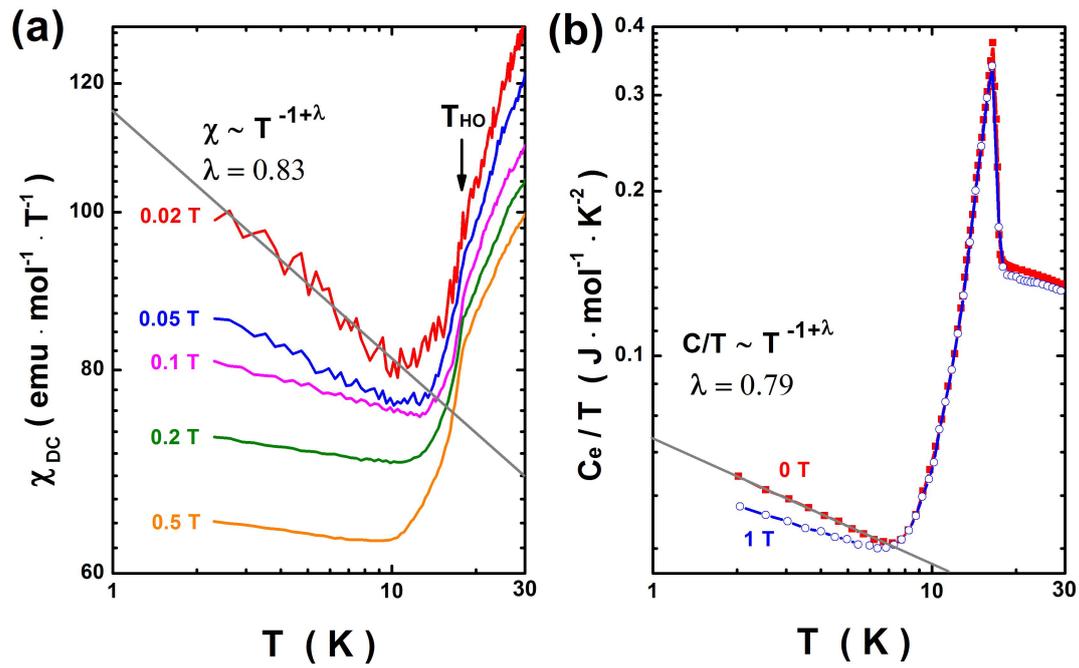

FIG 3. The fitted parameter, λ, of the Griffiths phase as a function of the magnetic field.

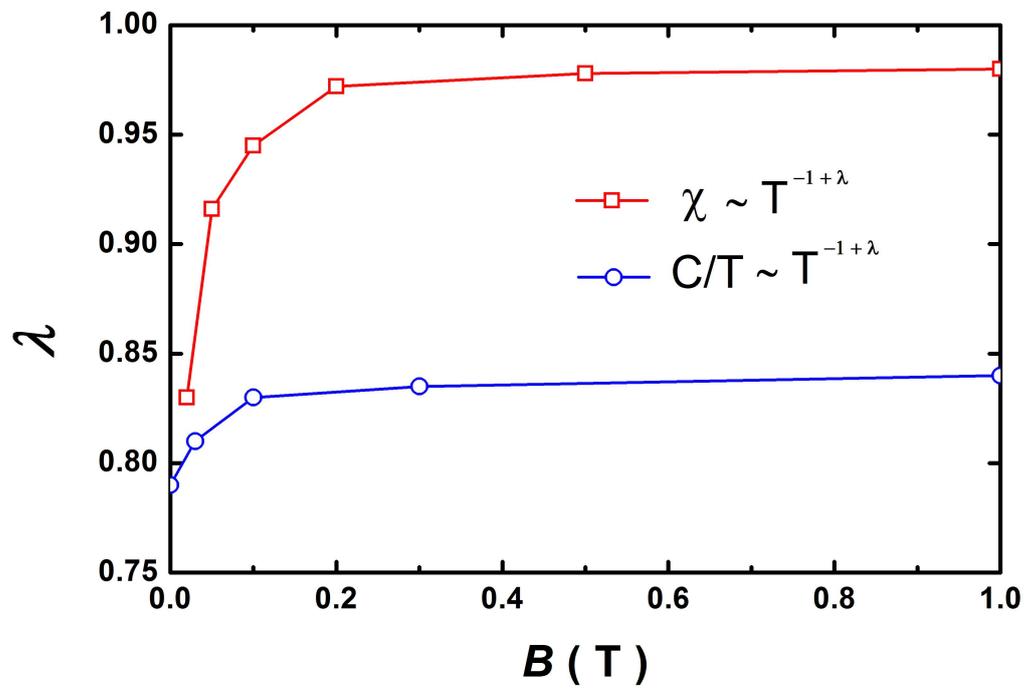

FIG. 4. Magnetic AC susceptibility measurements. (a) Frequency-dependent AC susceptibility measured as a function of the temperature. Frequency dispersion below $T_{HO}$ indicates the existence of short-range correlations, satisfying the Griffiths phase scenario. DC susceptibility is also plotted in the figure for comparison. (b) AC susceptibility measured at various DC bias magnetic fields. (c) The relative change of AC susceptibility, $[\chi_{AC}(B)- \chi_{AC}(0)] / \chi_{AC}(0) \times 100\%$, as a function of the magnetic field.

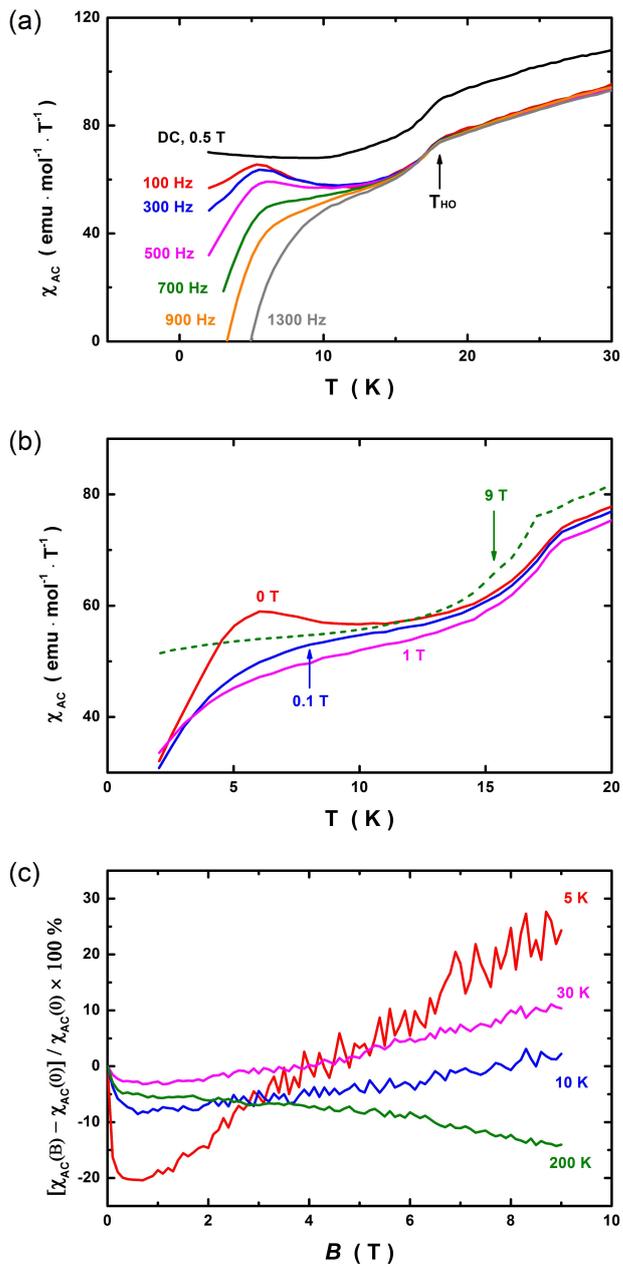

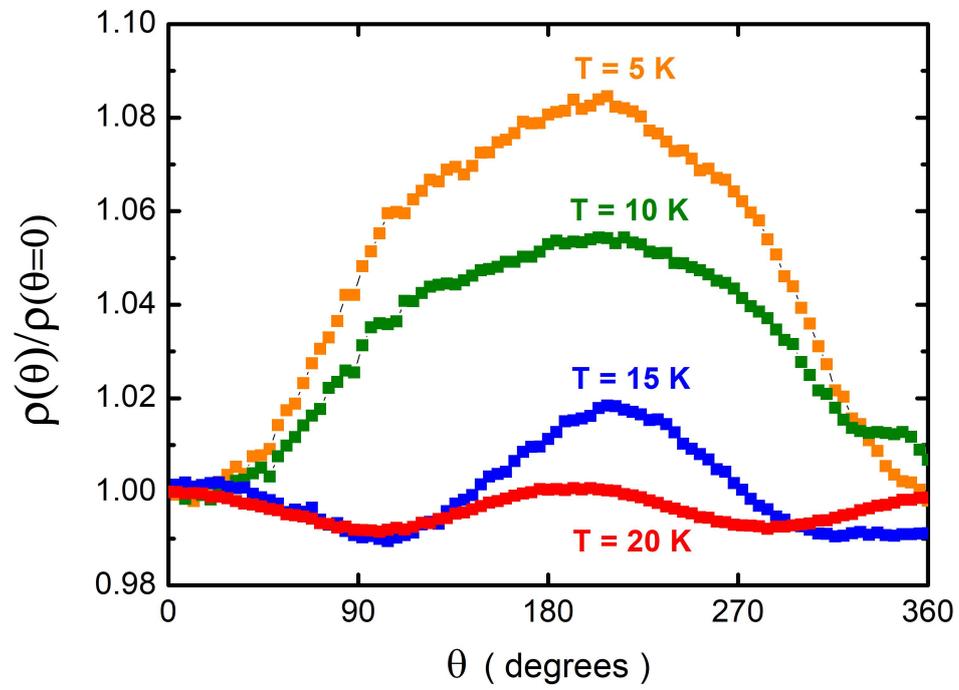

FIG. 5. The relative change of the resistivity, $\rho(\theta)/\rho(\theta=0)$, as a function of $\theta$ at 4 T.

# Supplementary materials for: Griffiths phase and symmetry breaking in the hidden-order phase of URu$_2$Si$_2$

Yi Liu, Wen Zhang, Xiaoying Wang, Donghua Xie, and Xinchun Lai

Science and Technology on Surface Physics and Chemistry Laboratory, Huafeng Xincun 9, Jiangyou 621908, China

In supplementary materials, we present the details of resistivity measurements in rotating magnetic fields. Fig. S1 shows the schematics of resistivity measurements. Fig. S2 shows the raw data of the resistivity measured at various temperatures.

[*] E-mail address: liuyifat@163.com

FIG. S1. Schematics of resistivity measurements of URu$_2$Si$_2$. In measurements, a standard four-wire AC technique was used with a PPMS rotator option. The AC current is 10 mA, operating at a frequency of 17 Hz. The current was always applied along the b-axis of the sample. The resistivity, $\rho$, was measured when the magnetic field of $\mu_0 H = 4$ T was rotated spanning the ac plane. In this geometry, $\rho$ is a periodic function of the azimuthal angle $\theta$ measured from the c axis. In the ac plane , the magnetic field with the rotator option has a high alignment precision (within 0.05°).

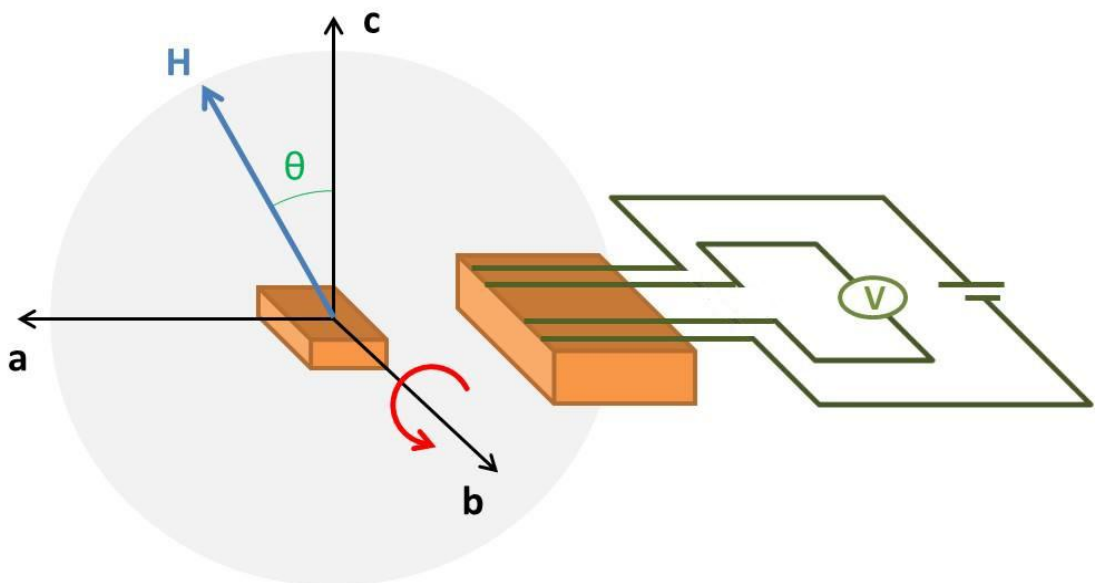

FIG. S2. Raw data of resistivity measurements at several temperatures, both above and below the hidden-order phase transition.

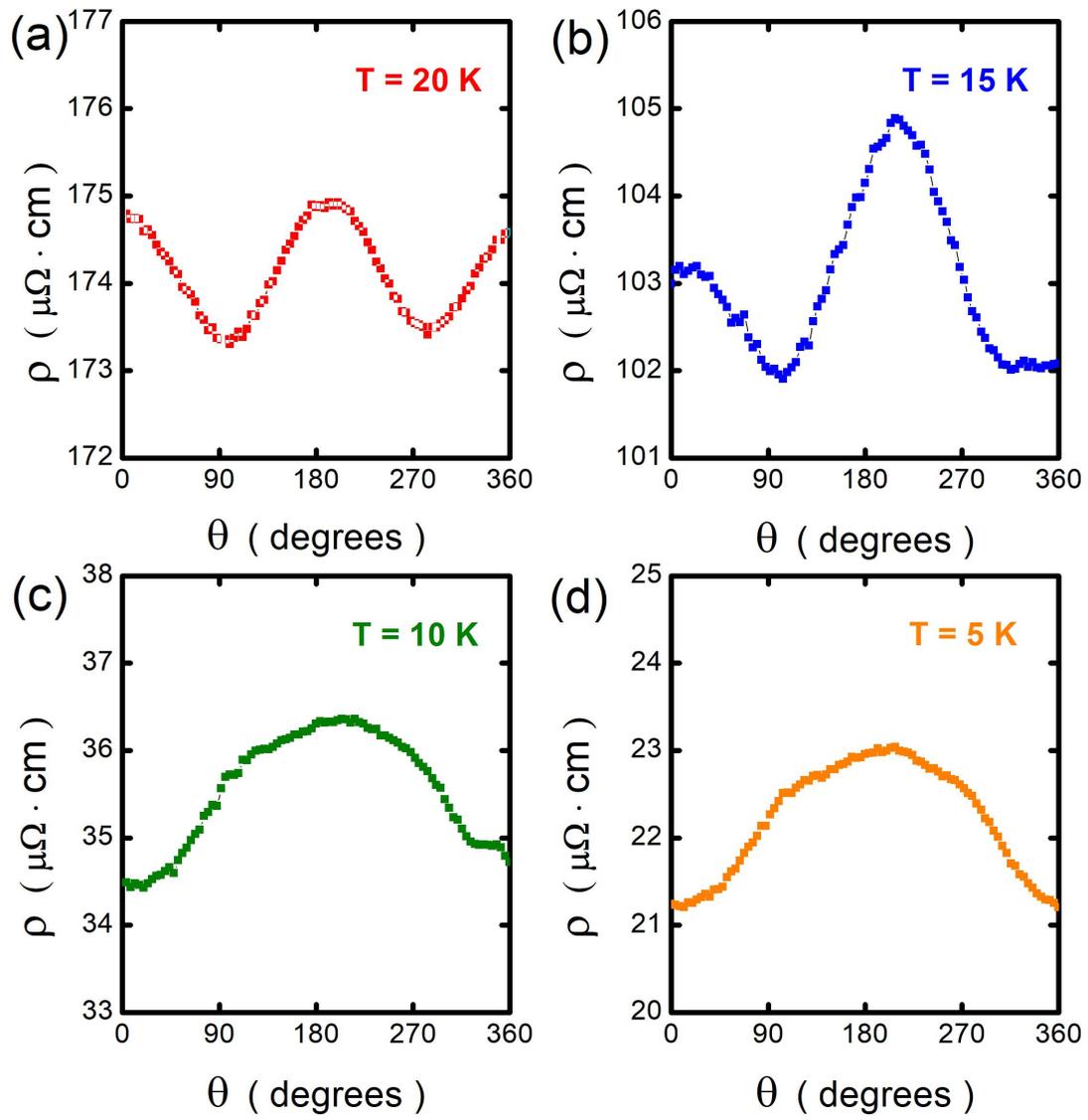